\documentclass[11pt]{article}
\usepackage[a4paper,margin=2.6cm]{geometry}
\usepackage{amsmath,amssymb,amsthm}
\usepackage{graphicx}
\usepackage{subcaption}
\usepackage{amsthm}

\usepackage{graphicx}
\usepackage{subcaption}
\usepackage{listings}  
\usepackage{xcolor} 
\usepackage{graphicx}
\usepackage{amsfonts}
\usepackage{amsmath}
\usepackage{url}
\usepackage{comment}
\usepackage{epstopdf}
\usepackage{hyperref} 
\usepackage{color}
\usepackage{rotating}
\usepackage[numbers,sort&compress]{natbib}
\usepackage{footnote}
\usepackage{bm}
\usepackage{hyperref}
\usepackage{float}
\usepackage{amsthm}

\usepackage{graphicx}
\usepackage{subcaption}
\usepackage{listings}  
\usepackage{xcolor} 
\newtheorem{theorem}{Theorem}
\newtheorem{proposition}{Proposition}
\newtheorem{corollary}{Corollary}
\theoremstyle{remark}
\newtheorem{remark}{Remark}

\def\ve{\varepsilon}
\def\llambda{{\tilde\lambda}}
\def\tsig{{\tilde\sigma}}
\newcommand{\av}[1]{\langle #1 \rangle}
\newcommand{\Tr}{\mathrm{Tr}}

\title{Critical step size of stochastic gradient descent\\ for a linear neuron:
an exact solution\\ and the role of the input spectrum}

\author{Sebastien Valeyre \thanks{Machina Capital, 21 rue de la Banque, 75002 Paris, France, svaleyre@machinacap.com} \thanks{Valeyre Research, 114 rue d'Antibes, 06400 Cannes,
France. E-mail: sv@valeyreresearch.com}}

\date{}

\begin{document}
\maketitle

\begin{abstract}
We study online stochastic gradient descent for a single linear neuron in a
teacher--student setting with Gaussian inputs of arbitrary covariance. The
mean-square errors of the $N$ weights obey a closed linear recursion of
dimension $N$, which we solve exactly. We prove that the dynamics converges if
and only if the step size is below a critical value $\lambda_c$, given by an
explicit scalar equation that involves all the eigenvalues of the input
covariance, and we obtain the stationary error of each weight, which diverges at
$\lambda_c$. This condition coincides with the classical mean-square stability
condition of the least-mean-squares filter; our derivation is elementary and
self-contained. We then show that $\lambda_c$ is controlled by the effective
rank $\Omega_1$ of the input covariance rather than by the number of parameters,
$1/(\Omega_1+2)\leq\lambda_c<1/\Omega_1$ in rescaled units: the critical step
size is inversely proportional to $N$ only for isotropic inputs, whereas it
decays more slowly, or saturates, for power-law spectra. Finally, we extend the
solution to mini-batches of size $B$: the critical step size grows linearly with
$B$ and saturates once $B$ exceeds the effective rank.
\end{abstract}


\section{Introduction}

Modern neural networks are trained by stochastic gradient descent (SGD) and its
variants, and the choice of the step size, or learning rate, is one of the most
important practical decisions \cite{Bottou18,Bengio12}. If the step size is too
small, learning is slow; if it is too large, the dynamics becomes unstable. As
the number of parameters $N$ of the models has grown by orders of magnitude
\cite{Geiger20,Martin19}, a natural question is how the largest admissible step
size depends on $N$.

In this paper we answer this question exactly in the simplest setting in which
it can be posed: a single linear neuron with $N$ inputs that learns online from
examples generated by a teacher neuron with the same architecture. The
teacher--student framework has a long history in the statistical mechanics of
learning \cite{Seung92,Watkin93,Engel01}, where online gradient descent was
analysed in detail \cite{Biehl95,SaadSolla95,Werfel05}. The same model is known
in signal processing as the least-mean-squares (LMS) adaptive filter of Widrow
and Hoff \cite{WidrowHoff60}. Its convergence in the mean square was established
for Gaussian inputs in \cite{Widrow76,Horowitz81,Feuer85} and is now textbook
material \cite{Haykin02,Sayed03}. In the machine-learning literature, SGD for
least-squares regression has been studied extensively
\cite{BachMoulines13,Defossez15,Jain18,Zou21}, and the largest step size was
related to the trace of the input covariance.

These three communities use different languages, and the results of one are not
always known to the others. In particular, the role of the input spectrum is
often overlooked when the question is phrased in terms of the number of
parameters: bounds based on the Lipschitz constant of the gradient involve only
the largest eigenvalue of the input covariance \cite{LeCun91,Bottou18}, which is
appropriate for deterministic gradient descent but not for its stochastic
version. Our purpose is to give a short, elementary and self-contained
derivation, in the language of the teacher--student model, and to make the
dependence on $N$ and on the spectrum explicit. Our contributions are the
following.
\begin{enumerate}
\item We show that the mean-square errors of the weights obey a closed linear
system of dimension $N$, whose matrix is a diagonal matrix plus a rank-one
matrix (Section~\ref{sec:dynamics}). No closure approximation is needed.
\item We prove that the dynamics converges if and only if the step size is
smaller than a critical value $\lambda_c$, the unique root of an explicit scalar
equation (Theorem~\ref{th:main}). This condition coincides with the
mean-square stability condition of the LMS filter with Gaussian inputs
\cite{Horowitz81,Feuer85}; the proof given here takes a few lines.
\item We give the stationary error of each weight in closed form, as well as
the convergence rate (Sections~\ref{sec:stationary} and \ref{sec:rate}).
\item We show that $\lambda_c$ is determined, up to a narrow interval, by the
effective rank of the input covariance and not by $N$
(Proposition~\ref{prop:bounds}). The scaling $\lambda_c\propto 1/N$ holds only
for isotropic inputs; for a power-law spectrum, $\lambda_c$ decays more slowly
with $N$ or tends to a constant (Section~\ref{sec:spectrum}).
\item We extend the solution to mini-batches of size $B$ and obtain the
critical step size in closed form in the isotropic case,
$\lambda_c(B)=B/(N+B+1)$, which exhibits the linear scaling with the batch size
used in practice \cite{Goyal17,Smith18} and its saturation
(Section~\ref{sec:batch}).
\end{enumerate}
The first three items revisit classical results with a different method; the
last two make explicit consequences that are, to our knowledge, not stated in
this form in the statistical-physics literature on learning. The limitations of
the model are discussed in Section~\ref{sec:discussion}.

\section{Model}
\label{sec:model}

The teacher is a linear neuron with $N$ fixed weights $\bar p_1,\ldots,\bar
p_N$, and the student is a linear neuron with weights $p_1(t),\ldots,p_N(t)$
that depend on the iteration $t$. At each iteration, both receive the same
random inputs $\ve_1(t),\ldots,\ve_N(t)$, and the output of the teacher is
corrupted by a noise $\ve_0(t)$. The error of the student is
\begin{equation}
\label{eq:R}
R(t) = \ve_0(t) + \sum\limits_{i=1}^N \bigl(p_i(t) - \bar{p}_i\bigr)\, \ve_i(t) .
\end{equation}
The student is trained by gradient descent on the instantaneous squared error,
with a constant step size $\llambda>0$:
\begin{equation}
\label{eq:sgd}
p_i(t+1) - p_i(t) = - \llambda\, \frac{\partial R^2(t)}{\partial p_i(t)}
 = - 2\llambda\, \ve_i(t)\, R(t) \qquad (i=1,\ldots,N).
\end{equation}
We assume that $\ve_0(t),\ve_1(t),\ldots,\ve_N(t)$ are independent Gaussian
variables with mean zero and standard deviations $\sigma_0,\tsig_1,\ldots,
\tsig_N$, drawn independently at each iteration (online learning: each example
is used once). As a consequence, the weights $p_i(t)$, which depend only on
the inputs at times $t'<t$, are independent of $\ve_j(t)$. Without loss of
generality, the standard deviations are ordered and strictly positive,
\begin{equation}
\tsig_1 \geq \tsig_2 \geq \ldots \geq \tsig_N > 0 ,
\end{equation}
since a weight whose input has zero variance does not evolve.

\begin{remark}
\label{rem:corr}
The assumption of independent inputs is not restrictive. If the inputs are
jointly Gaussian with a covariance matrix $C$, one can rotate the inputs and
the weights to the eigenbasis of $C$: Eq.~(\ref{eq:sgd}) keeps the same form,
the rotated inputs are independent, and the total weight error is invariant.
The quantities $\tsig_i^2$ are then the eigenvalues of $C$.
\end{remark}

It is convenient to rescale the variances and the step size by the largest
variance $\sigma^2=\tsig_1^2$:
\begin{equation}
\label{eq:rescaling}
\sigma_i = \tsig_i/\sigma \in (0,1], \qquad
\lambda = \llambda\, \sigma^2 , \qquad
\nu = \sigma_0^2/\sigma^2 .
\end{equation}
We characterise the convergence of the student by the mean-square error of each
weight and by their weighted sums,
\begin{equation}
\label{eq:Fn}
u_i(t) = \av{\bigl(p_i(t) - \bar{p}_i\bigr)^2} , \qquad
F_n(t) = \sum\limits_{i=1}^N \sigma_i^{2n}\, u_i(t) ,
\end{equation}
where $\av{\cdots}$ denotes the expectation over the inputs and the noise. The
total weight error is $F_0(t)$, and the excess prediction error of the student
is $\av{R^2(t)}-\sigma_0^2=\sigma^2 F_1(t)$. We also set
\begin{equation}
\label{eq:Omega}
\Omega_n = \sum\limits_{i=1}^N \sigma_i^{2n} , \qquad 1\leq \Omega_n\leq N .
\end{equation}
The quantity $\Omega_1=\Tr\, C/\|C\|$ is the effective rank of the input
covariance: it is equal to $N$ for isotropic inputs and can be much smaller
than $N$ when a few directions dominate.

\section{Exact dynamics of the weight errors}
\label{sec:dynamics}

\begin{proposition}
\label{prop:closed}
The mean-square errors of the weights obey the closed linear recursion
\begin{equation}
\label{eq:u_rec}
u_i(t+1) = \rho_i\, u_i(t) + 4\lambda^2 \sigma_i^2 \bigl[F_1(t) + \nu\bigr] ,
\qquad
\rho_i = 1 - 4\lambda \sigma_i^2 + 8\lambda^2\sigma_i^4 .
\end{equation}
\end{proposition}

\begin{proof}
Let $d_i(t)=p_i(t)-\bar p_i$. From Eq.~(\ref{eq:sgd}),
\begin{equation}
u_i(t+1) = u_i(t) - 4\llambda\, \av{d_i(t)\,\ve_i(t)\,R(t)}
 + 4\llambda^2\, \av{\ve_i^2(t)\, R^2(t)} .
\end{equation}
Since $d_j(t)$ is independent of $\ve_k(t)$ and
$\av{\ve_i(t)\ve_j(t)}=\tsig_i^2\delta_{i,j}$, the second term is
\begin{equation}
\av{d_i\,\ve_i\,R} = \sum\limits_{j=1}^N \av{d_i d_j}\,\av{\ve_i\ve_j}
 = \tsig_i^2\, u_i(t) .
\end{equation}
For the third term, the contributions that are odd in $\ve_0$ or in $\ve_j$
vanish, so that
\begin{equation}
\av{\ve_i^2 R^2} = \sigma_0^2\tsig_i^2
 + \sum\limits_{j=1}^N \av{\ve_i^2\ve_j^2}\, u_j(t) .
\end{equation}
For Gaussian variables, $\av{\ve_i^2\ve_j^2}=\tsig_i^2\tsig_j^2
+2\tsig_i^4\delta_{i,j}$, from which
\begin{equation}
\label{eq:S2}
\av{\ve_i^2 R^2} = \sigma_0^2\tsig_i^2 + 2\tsig_i^4\, u_i(t)
 + \tsig_i^2 \sum\limits_{j=1}^N \tsig_j^2\, u_j(t) .
\end{equation}
Combining these results and using the rescaling (\ref{eq:rescaling}) yields
Eq.~(\ref{eq:u_rec}).
\end{proof}

Two features of Eq.~(\ref{eq:u_rec}) deserve a comment. First, the
cross-correlations $\av{d_i d_j}$ with $i\ne j$ do not appear: the $N$
mean-square errors form a closed system. Second, the weights are coupled only
through the excess prediction error $F_1(t)$, i.e., through a single scalar. In
matrix form, with ${\bf u}(t)=(u_1(t),\ldots,u_N(t))^\dagger$ and
${\bf s}=(\sigma_1^2,\ldots,\sigma_N^2)^\dagger$, we have
\begin{equation}
\label{eq:matrix}
{\bf u}(t+1) = M\, {\bf u}(t) + 4\lambda^2\nu\, {\bf s} , \qquad
M = D + 4\lambda^2\, {\bf s}\,{\bf s}^\dagger , \qquad
D = \mathrm{diag}(\rho_1,\ldots,\rho_N) .
\end{equation}
The matrix $M$ is symmetric, and it is the sum of a diagonal matrix and of a
rank-one matrix. Since $\rho_i=(1-2\lambda\sigma_i^2)^2+4\lambda^2\sigma_i^4>0$,
it is also positive definite, with strictly positive entries. Whenever $I-M$ is
invertible, the solution of Eq.~(\ref{eq:matrix}) is
\begin{equation}
\label{eq:solution}
{\bf u}(t) = {\bf u}(\infty) + M^t \bigl[{\bf u}(0) - {\bf u}(\infty)\bigr] ,
\qquad {\bf u}(\infty) = 4\lambda^2\nu\, (I-M)^{-1}{\bf s} .
\end{equation}
Multiplying Eq.~(\ref{eq:u_rec}) by $\sigma_i^{2n}$ and summing over $i$ gives
a hierarchy of equations for the moments $F_n(t)$, which is discussed in
\ref{app:hierarchy}.

\begin{remark}
The mean errors obey $\av{d_i(t+1)}=(1-2\lambda\sigma_i^2)\av{d_i(t)}$ and
vanish exponentially for any $\lambda<1$. Convergence in the mean is therefore a
much weaker requirement than convergence in the mean square, which is studied
below.
\end{remark}

\section{Critical step size}
\label{sec:critical}

For $0\leq\lambda<1/2$, we define
\begin{equation}
\label{eq:Sigma}
\Sigma_n(\lambda) = \sum\limits_{i=1}^N \frac{\sigma_i^{2n}}{1-2\lambda \sigma_i^2} .
\end{equation}
The function $\lambda\,\Sigma_1(\lambda)$ is continuous and strictly increasing
on $(0,1/2)$, vanishes at $\lambda=0$ and diverges as $\lambda\to 1/2$ (because
$\sigma_1=1$). There exists therefore a unique $\lambda_c\in(0,1/2)$ such that
\begin{equation}
\label{eq:lambda_c}
\lambda_c\, \Sigma_1(\lambda_c) = 1 .
\end{equation}

\begin{theorem}
\label{th:main}
{\rm (i)} If $\lambda<\lambda_c$, all the eigenvalues of $M$ lie in $(0,1)$,
and ${\bf u}(t)$ converges exponentially to ${\bf u}(\infty)$ for any initial
condition. {\rm (ii)} If $\lambda>\lambda_c$, the largest eigenvalue of $M$
exceeds $1$, and the total error $F_0(t)$ diverges exponentially, except in the
trivial case $\nu=0$ and ${\bf u}(0)=0$.
\end{theorem}

\begin{proof}
Since $M$ is symmetric and positive definite, its eigenvalues are real and
positive, and they are all smaller than $1$ if and only if $I-M=(I-D)-4\lambda^2
{\bf s}{\bf s}^\dagger$ is positive definite. If $\rho_i\geq 1$ for some $i$,
the $i$-th diagonal element of $I-M$ is negative and $I-M$ is not positive
definite. Otherwise $A=I-D$ is positive definite, and $A-4\lambda^2{\bf s}{\bf
s}^\dagger$ is positive definite if and only if $4\lambda^2\,{\bf
s}^\dagger A^{-1}{\bf s}<1$ (indeed, $A^{-1/2}(A-4\lambda^2{\bf s}{\bf
s}^\dagger)A^{-1/2}$ is the identity minus a rank-one matrix, whose eigenvalues
are $1$ and $1-4\lambda^2{\bf s}^\dagger A^{-1}{\bf s}$). Now $1-\rho_i=4\lambda\sigma_i^2
(1-2\lambda\sigma_i^2)$, so that $\rho_i<1$ for all $i$ if and only if
$\lambda<1/2$, and
\begin{equation}
4\lambda^2\,{\bf s}^\dagger(I-D)^{-1}{\bf s}
 = \sum\limits_{i=1}^N \frac{4\lambda^2\sigma_i^4}{4\lambda\sigma_i^2(1-2\lambda\sigma_i^2)}
 = \lambda\,\Sigma_1(\lambda) .
\end{equation}
The eigenvalues of $M$ are thus all smaller than $1$ if and only if
$\lambda<1/2$ and $\lambda\Sigma_1(\lambda)<1$, i.e., if and only if
$\lambda<\lambda_c$. Statement (i) then follows from Eq.~(\ref{eq:solution}).
For (ii), we first show that the largest eigenvalue $\mu_{\max}$ of $M$
exceeds $1$ when $\lambda>\lambda_c$. If $\lambda\geq 1/2$, then $\rho_1\geq 1$
and $\mu_{\max}\geq M_{1,1}=\rho_1+4\lambda^2>1$. If $\lambda_c<\lambda<1/2$,
then $a={\bf s}^\dagger(I-D)^{-1}{\bf s}$ is such that $4\lambda^2 a=
\lambda\Sigma_1(\lambda)>1$, and the vector ${\bf x}=(I-D)^{-1}{\bf s}$ gives
${\bf x}^\dagger(I-M){\bf x}=a\,(1-4\lambda^2 a)<0$, so that $I-M$ has a
negative eigenvalue. Since $M$ has strictly positive entries, the eigenvector
${\bf v}$ associated with $\mu_{\max}$ can be chosen with strictly positive
components (Perron--Frobenius theorem). Projecting Eq.~(\ref{eq:matrix}) onto
${\bf v}$ gives ${\bf v}^\dagger{\bf u}(t+1)=\mu_{\max}\,{\bf v}^\dagger{\bf
u}(t)+4\lambda^2\nu\,{\bf v}^\dagger{\bf s}$. As $u_i(t)\geq 0$, the quantity
${\bf v}^\dagger{\bf u}(t)$ is positive as soon as $\nu>0$ or ${\bf u}(0)\ne
0$, and it grows as $\mu_{\max}^t$.
\end{proof}

At $\lambda=\lambda_c$ the largest eigenvalue is equal to $1$, and the error
grows linearly in time if $\nu>0$. In the original units, the critical step
size is $\llambda_c=\lambda_c/\tsig_1^2$.

\begin{corollary}[Isotropic inputs]
\label{cor:iso}
If all the variances are equal, $\sigma_i=1$, then $\Sigma_1(\lambda)=
N/(1-2\lambda)$ and
\begin{equation}
\label{eq:cond_equal}
\lambda_c = \frac{1}{N+2} .
\end{equation}
\end{corollary}

In the isotropic case the critical step size is thus inversely proportional to
the number of parameters for large $N$. The next result shows that, in general,
the relevant quantity is not $N$ but the effective rank $\Omega_1$.

\begin{proposition}[Bounds]
\label{prop:bounds}
For any set of variances,
\begin{equation}
\label{eq:bounds}
\frac{1}{\Omega_1+2} \;\leq\; \lambda_c \;\leq\; \frac{1}{\Omega_1+2\sigma_N^2}
\;<\; \frac{1}{\Omega_1} ,
\end{equation}
with equality in the first two inequalities if and only if the inputs are
isotropic.
\end{proposition}

\begin{proof}
Since $\sigma_N^2\leq\sigma_i^2\leq 1$, one has $(1-2\lambda\sigma_N^2)^{-1}\leq
(1-2\lambda\sigma_i^2)^{-1}\leq(1-2\lambda)^{-1}$, hence
$\lambda\Omega_1/(1-2\lambda\sigma_N^2)\leq\lambda\Sigma_1(\lambda)\leq
\lambda\Omega_1/(1-2\lambda)$. Setting $\lambda=\lambda_c$ and using
Eq.~(\ref{eq:lambda_c}) gives Eq.~(\ref{eq:bounds}).
\end{proof}

Since $\Omega_1\leq N$, Eq.~(\ref{eq:bounds}) implies $\lambda_c\geq 1/(N+2)$:
among all input distributions with a given largest variance, the isotropic one
is the most constraining. In the original units, Eq.~(\ref{eq:bounds}) reads
\begin{equation}
\label{eq:bounds_orig}
\frac{1}{\Tr\, C+2\|C\|} \;\leq\; \llambda_c \;<\; \frac{1}{\Tr\, C} ,
\end{equation}
where $\Tr\,C=\sum_i\tsig_i^2$ is the mean squared norm of the input and
$\|C\|=\tsig_1^2$ its largest variance. The two bounds differ by a relative
amount $2/\Omega_1$ at most, so that $\llambda_c\simeq 1/\Tr\,C$ as soon as
the effective rank is large. By contrast, deterministic gradient descent on the
average error $\av{R^2}$ is stable for $\llambda<1/\|C\|$, a condition that
involves only the largest eigenvalue \cite{LeCun91}. The stochastic dynamics is
therefore more constrained than the deterministic one by a factor of order
$\Omega_1$.

\paragraph{Relation to the LMS filter.}
Equation~(\ref{eq:sgd}) is the LMS recursion ${\bf w}(t+1)={\bf w}(t)+\mu\,
e(t)\,{\bf x}(t)$ with $\mu=2\llambda$. Writing $a_i=\tsig_i^2$ for the
eigenvalues of the input covariance, the condition $\lambda<\lambda_c$ reads
\begin{equation}
\mu < \frac{1}{a_1} \qquad \text{and} \qquad
\sum\limits_{i=1}^N \frac{\mu\, a_i}{2-2\mu\, a_i} < 1 ,
\end{equation}
which is the mean-square stability condition obtained for the LMS filter with
real Gaussian inputs \cite{Horowitz81,Feuer85}, and the upper bound in
Eq.~(\ref{eq:bounds_orig}) is the practical rule $\mu<2/\Tr\,C$ of the
adaptive-filtering literature \cite{Widrow76,Haykin02}. The present derivation
recovers these results from the teacher--student formulation.

\begin{remark}[Non-Gaussian inputs]
\label{rem:kurtosis}
The term $2\tsig_i^4$ in Eq.~(\ref{eq:S2}) comes from the fourth moment of the
Gaussian distribution. If the inputs are independent with a common kurtosis
$\kappa=\av{\ve_i^4}/\tsig_i^4$, it is replaced by $(\kappa-1)\tsig_i^4$. In
the isotropic case one finds $\lambda_c=1/(N+\kappa-1)$, which reduces to
Eq.~(\ref{eq:cond_equal}) for $\kappa=3$ and gives $\lambda_c=1/N$ for binary
inputs ($\kappa=1$). The correction to the leading behaviour $1/N$ is thus a
signature of the tails of the input distribution.
\end{remark}

\section{Stationary errors}
\label{sec:stationary}

For $\lambda<\lambda_c$, setting $u_i(t+1)=u_i(t)=u_i(\infty)$ in
Eq.~(\ref{eq:u_rec}) gives $u_i(\infty)=\lambda\,[F_1(\infty)+\nu]/
(1-2\lambda\sigma_i^2)$. Multiplying by $\sigma_i^2$ and summing over $i$
determines $F_1(\infty)$, from which
\begin{equation}
\label{eq:u_inf}
u_i(\infty) = \frac{\nu\,\lambda}{\bigl(1-2\lambda\sigma_i^2\bigr)
 \bigl(1-\lambda\,\Sigma_1(\lambda)\bigr)} , \qquad
F_n(\infty) = \frac{\nu\,\lambda\,\Sigma_n(\lambda)}{1-\lambda\,\Sigma_1(\lambda)} .
\end{equation}
The second relation holds for any real $n$, in particular for the total weight
error ($n=0$) and for the excess prediction error ($n=1$). The ratio of the
stationary excess prediction error to the noise variance, called the
misadjustment in the adaptive-filtering literature \cite{Widrow76}, is
\begin{equation}
\label{eq:misadj}
\frac{\sigma^2 F_1(\infty)}{\sigma_0^2}
 = \frac{\lambda\,\Sigma_1(\lambda)}{1-\lambda\,\Sigma_1(\lambda)} .
\end{equation}
Three consequences follow. (a) Without noise ($\nu=0$) the student recovers the
teacher exactly. (b) With noise, the stationary error is proportional to the
noise variance and, for small step sizes, to the step size: $u_i(\infty)\simeq
\nu\lambda$, i.e., $u_i(\infty)\simeq\llambda\,\sigma_0^2$ in the original
units, the same for all the weights. (c) All the stationary errors diverge as
$(\lambda_c-\lambda)^{-1}$ when the step size approaches the critical value:
$\lambda_c$ is not only the limit of stability but also the point at which the
accuracy of the student is lost.

Using the identity $2\lambda\,\Sigma_1(\lambda)=\Sigma_0(\lambda)-N$, the total
weight error can also be written as $F_0(\infty)=2\nu\lambda\,
\Sigma_0(\lambda)/[N+2-\Sigma_0(\lambda)]$, and the stability condition as
$\Sigma_0(\lambda)<N+2$. Since $\Sigma_0(\lambda)$ increases with $\lambda$,
$F_0(\infty)$ is an increasing function of the step size. In the isotropic
case, $F_0(\infty)=\nu\lambda N/[1-\lambda(N+2)]$.

\begin{figure}[H]
	\centering
	\begin{subfigure}[t]{0.48\textwidth}
		\centering
		\includegraphics[width=\textwidth]{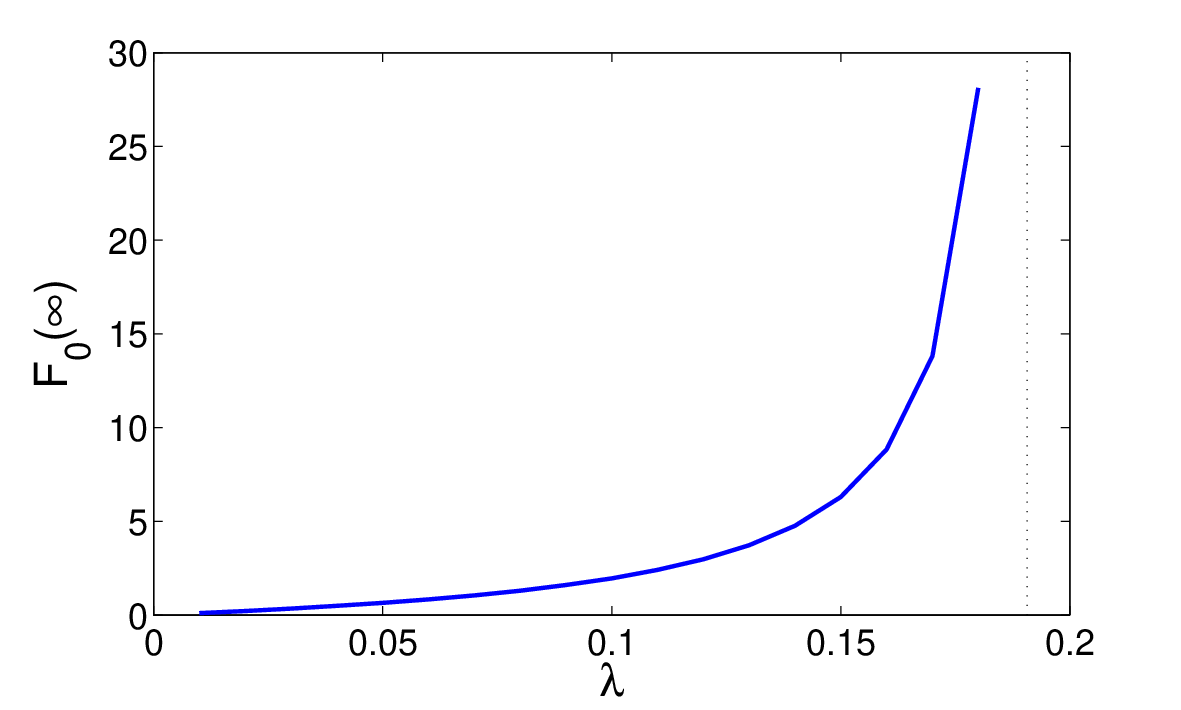}
		\caption{Stationary weight error $F_0(\infty)$ as a function of the
			rescaled step size $\lambda$: Eq.~(\ref{eq:u_inf}) (line) and Monte Carlo
			simulations of Eq.~(\ref{eq:sgd}) (circles). The vertical dotted line indicates
			$\lambda_c\simeq 0.1906$.}
		\label{fig}
	\end{subfigure}
	\hfill
	\begin{subfigure}[t]{0.48\textwidth}
		\centering
		\includegraphics[width=\textwidth]{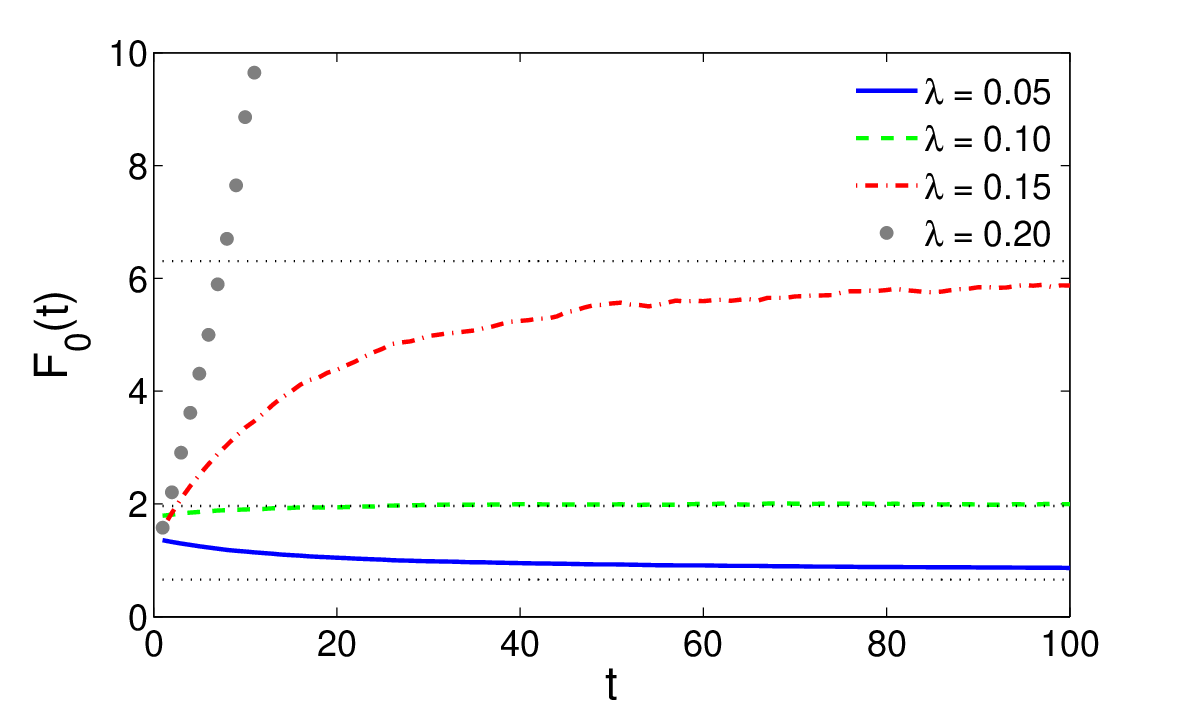}
		\caption{ Weight error $F_0(t)$ starting from $u_i(0)=1$,
			for four step sizes: Eq.~(\ref{eq:solution}) (lines) and simulations averaged
			over $4\times10^4$ realisations (symbols). The dynamics diverges for
			$\lambda=0.20>\lambda_c$.}
		\label{fig}
	\end{subfigure}
	\caption{Single linear neuron with $N=10$ inputs, $\sigma_i=1-(i-1)/N$ and
		$\nu=1$}
	\label{fig:F0}
\end{figure}

Figure~\ref{fig:F0} compares these predictions with Monte Carlo simulations of
Eq.~(\ref{eq:sgd}) for $N=10$ and a linear profile of standard deviations,
$\sigma_i=1-(i-1)/N$, for which $\Omega_1=3.85$ and $\lambda_c\simeq 0.1906$,
whereas $1/(N+2)\simeq 0.083$.

\section{Convergence rate}
\label{sec:rate}

According to Eq.~(\ref{eq:solution}), the approach to the stationary state is
governed at long times by the largest eigenvalue $\mu_{\max}$ of $M$:
$F_n(t)-F_n(\infty)\propto\mu_{\max}^t$. Since $M$ is a rank-one perturbation
of a diagonal matrix, its eigenvalues that differ from the $\rho_i$ are the
roots of a secular equation \cite{Golub73}, and $\mu_{\max}$ is the unique
solution larger than $\max_i\rho_i$ of
\begin{equation}
\label{eq:secular}
4\lambda^2 \sum\limits_{i=1}^N \frac{\sigma_i^4}{\mu-\rho_i} = 1 .
\end{equation}
Setting $\mu=1$ in the left-hand side gives back $\lambda\,\Sigma_1(\lambda)$,
in agreement with Theorem~\ref{th:main}. In the isotropic case,
\begin{equation}
\label{eq:q}
\mu_{\max} = 1 - 4\lambda + 4\lambda^2 (N+2) ,
\end{equation}
which is minimal for $\lambda=\lambda_c/2=1/(2(N+2))$, where
$\mu_{\max}=1-1/(N+2)$: the fastest convergence takes of the order of $N$
iterations, i.e., of $N$ examples, and is obtained for half of the critical
step size.

In the general case, $\mu_{\max}>\rho_i$ for all $i$ implies $1-\mu_{\max}<
4\lambda\sigma_N^2$, and since $\lambda<\lambda_c<1/\Omega_1$, the relaxation
time $\tau=1/(1-\mu_{\max})$ obeys
\begin{equation}
\label{eq:tau}
\tau > \frac{\Omega_1}{4\,\sigma_N^2} = \frac{\Tr\,C}{4\,\tsig_N^2} .
\end{equation}
The weights associated with the inputs of small variance are learnt slowly, and
the step size cannot be increased to compensate because it is limited by the
sum of all the variances. A small step size reduces the stationary error
(\ref{eq:u_inf}) but increases the relaxation time, so that for a finite number
of iterations there is an optimal step size, as illustrated in
Fig.~\ref{fig:F0}(b), where the curves for $\lambda=0.05$ and $\lambda=0.10$
cross.

\section{Dependence on the input spectrum}
\label{sec:spectrum}

\begin{figure}[t]
\begin{center}
\includegraphics[width=0.98\textwidth]{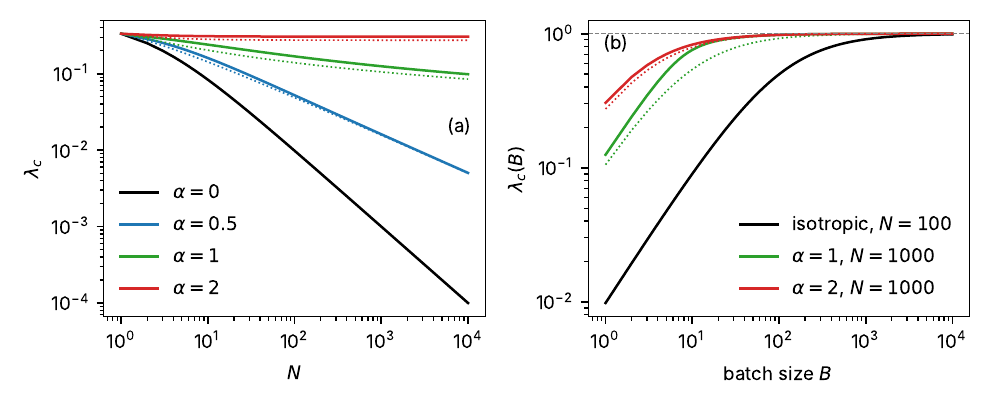}
\end{center}
\caption{(a) Critical step size $\lambda_c$ as a function of the number of
inputs $N$ for a power-law spectrum $\sigma_i^2=i^{-\alpha}$ (solid lines),
and the lower bound $1/(\Omega_1+2)$ of Eq.~(\ref{eq:bounds}) (dotted lines);
the two coincide for $\alpha=0$. (b) Critical step size $\lambda_c(B)$ as a
function of the batch size $B$ (solid lines), and the lower bound
$B/(\Omega_1+B+1)$ of Eq.~(\ref{eq:bounds_B}) (dotted lines), which is exact in
the isotropic case. The horizontal dashed line is the limit of deterministic
gradient descent, $\lambda=1$.}
\label{fig:spectrum}
\end{figure}

Proposition~\ref{prop:bounds} shows that the scaling of the critical step size
with the number of parameters is entirely determined by the growth of the
effective rank $\Omega_1$ with $N$. To illustrate this point, consider a
power-law spectrum,
\begin{equation}
\sigma_i^2 = i^{-\alpha} \qquad (i=1,\ldots,N),
\end{equation}
which is a common description of natural data \cite{Ruderman94} and of the
kernels of wide networks \cite{Bordelon20}. Then $\Omega_1=\sum_{i=1}^N
i^{-\alpha}$, and Eq.~(\ref{eq:bounds}) gives three regimes for large $N$, to leading
order:
\begin{equation}
\label{eq:regimes}
\lambda_c \simeq
\begin{cases}
(1-\alpha)\, N^{-(1-\alpha)} & (0\leq\alpha<1), \\[1mm]
1/\ln N & (\alpha=1), \\[1mm]
\text{constant} & (\alpha>1) ,
\end{cases}
\end{equation}
where the constant lies between $1/(\zeta(\alpha)+2)$ and $1/\zeta(\alpha)$,
$\zeta$ being the Riemann zeta function. The scaling $\lambda_c\propto 1/N$ is
recovered only for $\alpha=0$. For $\alpha=2$, for instance, the numerical
solution of Eq.~(\ref{eq:lambda_c}) gives $\lambda_c\simeq 0.306$ for all
$N\geq 100$, whereas $\lambda_c\simeq 0.125$ for $\alpha=1$ and $N=1000$
(Fig.~\ref{fig:spectrum}(a)). In other words, adding inputs of small variance
does not reduce the critical step size. It does, however, increase the
relaxation time, which by Eq.~(\ref{eq:tau}) is larger than
$\Omega_1 N^{\alpha}/4$.

\section{Mini-batches}
\label{sec:batch}

In practice, the gradient is averaged over a mini-batch of $B$ independent
examples:
\begin{equation}
\label{eq:sgd_B}
p_i(t+1) - p_i(t) = - \frac{2\llambda}{B} \sum\limits_{b=1}^B
 \ve_i^{(b)}(t)\, R^{(b)}(t) .
\end{equation}
The derivation of Section~\ref{sec:dynamics} is easily adapted. The $B$ terms
with the same example contribute as in Eq.~(\ref{eq:S2}), and the $B(B-1)$
terms with two different examples contribute $\tsig_i^4\,u_i(t)$ each, because
$\av{\ve_i^{(b)}R^{(b)}}=\tsig_i^2\,d_i$ for given weights. One obtains
\begin{equation}
\label{eq:u_rec_B}
u_i(t+1) = \rho_i^{(B)}\, u_i(t) + \frac{4\lambda^2}{B}\, \sigma_i^2
 \bigl[F_1(t) + \nu\bigr] , \qquad
\rho_i^{(B)} = 1 - 4\lambda \sigma_i^2
 + 4\Bigl(1+\frac{1}{B}\Bigr)\lambda^2\sigma_i^4 ,
\end{equation}
which has the same structure as Eq.~(\ref{eq:u_rec}) and reduces to it for
$B=1$. One checks that $\rho_i^{(B)}>0$ and that $1-\rho_i^{(B)}=4\lambda
\sigma_i^2\,[1-(1+1/B)\lambda\sigma_i^2]$, so that the proof of
Theorem~\ref{th:main} applies with $4\lambda^2$ replaced by $4\lambda^2/B$ and
gives the following result.

\begin{theorem}
\label{th:batch}
The dynamics (\ref{eq:sgd_B}) converges in the mean square if and only if
$\lambda<\lambda_c(B)$, where $\lambda_c(B)$ is the unique solution in
$(0,B/(B+1))$ of
\begin{equation}
\label{eq:lambda_c_B}
\sum\limits_{i=1}^N \frac{\lambda\,\sigma_i^2}{B-(B+1)\lambda\,\sigma_i^2} = 1 .
\end{equation}
It obeys
\begin{equation}
\label{eq:bounds_B}
\frac{B}{\Omega_1+B+1} \;\leq\; \lambda_c(B) \;\leq\;
\frac{B}{\Omega_1+(B+1)\sigma_N^2} ,
\end{equation}
and, in the isotropic case,
\begin{equation}
\label{eq:iso_B}
\lambda_c(B) = \frac{B}{N+B+1} .
\end{equation}
\end{theorem}

The stationary errors are obtained as in Section~\ref{sec:stationary}:
$u_i(\infty)=\nu\lambda/\{[B-(B+1)\lambda\sigma_i^2]\,[1-\lambda
\Sigma_1^{(B)}(\lambda)]\}$, where $\lambda\Sigma_1^{(B)}(\lambda)$ denotes the
left-hand side of Eq.~(\ref{eq:lambda_c_B}).

Equations (\ref{eq:bounds_B}) and (\ref{eq:iso_B}) describe two regimes
(Fig.~\ref{fig:spectrum}(b)). As long as the batch size is small compared with
the effective rank, $B\ll\Omega_1$, the critical step size is proportional to
the batch size, $\lambda_c(B)\simeq B/\Omega_1$. This is the linear scaling
rule used to train large networks \cite{Goyal17,Smith18}, and it follows here
from the fact that the noise of the gradient is reduced by a factor $B$. When
$B\gg\Omega_1$, the critical step size saturates at $\lambda_c=1$, i.e.,
$\llambda_c=1/\|C\|$, which is the stability limit of deterministic gradient
descent. The crossover takes place at a batch size of the order of the
effective rank, in agreement with the analysis of mini-batch SGD for
least-squares regression in \cite{Jain18}.

In the isotropic case, the largest eigenvalue is $\mu_{\max}=1-4\lambda
+4\lambda^2(N+B+1)/B$, which is minimal for $\lambda=\lambda_c(B)/2$, where
$1-\mu_{\max}=B/(N+B+1)$. The convergence rate per example,
$(1-\mu_{\max})/B=1/(N+B+1)$, decreases with $B$: in this model, a larger batch
allows a larger step size but does not reduce the number of examples needed to
learn.

\section{Discussion and conclusion}
\label{sec:discussion}

We have solved exactly the online learning dynamics of a linear neuron trained
by stochastic gradient descent in a teacher--student setting. The mean-square
errors of the weights form a closed linear system of dimension $N$ whose
stability is controlled by a single scalar function of the step size. The
dynamics converges if and only if the step size is below the critical value
defined by Eq.~(\ref{eq:lambda_c}), at which the stationary error diverges.

The main message concerns the dependence on the number of parameters. The
statement that the largest admissible step size is inversely proportional to
$N$ is correct for isotropic inputs, for which $\lambda_c=1/(N+2)$. In general,
however, the critical step size is governed by the effective rank of the input
covariance, $\llambda_c\simeq 1/\Tr\,C$, and it becomes independent of $N$ when
the spectrum decays fast enough. What a large number of weakly excited
directions does affect is the time needed to learn them. With mini-batches, the
critical step size increases linearly with the batch size up to a batch size of
the order of the effective rank.

Several of these results are known in other forms. The stability condition and
the misadjustment are those of the LMS filter
\cite{Horowitz81,Feuer85,Haykin02}, and the role of the trace of the covariance
and of the batch size has been established for least-squares regression under
more general assumptions on the inputs \cite{BachMoulines13,Jain18,Zou21}. The
interest of the present approach is its simplicity: the Gaussian
teacher--student model gives closed formulas for the threshold, the stationary
error of each weight and the convergence rate, from which the scalings with
$N$, with the spectrum and with the batch size can be read directly.

The model has clear limitations. A single linear neuron is equivalent to linear
regression: there is no hidden layer, no feature learning and a single
minimum. Each example is used once, so that the results describe the online
regime and not the repeated use of a finite training set. For deep nonlinear
networks, the curvature of the loss changes during training, and gradient
descent is observed to operate close to the edge of stability
\cite{Cohen21,Lewkowycz20,Wu18}; the dependence of the learning rate on the
width is then tied to the parametrisation of the network \cite{Jacot18,Yang21}.
The exact solution presented here can serve as a reference case for these
studies: in the regime where a wide network behaves as a linear model in its
parameters, the quantities $\tsig_i^2$ are the eigenvalues of the corresponding
kernel, and the critical step size is set by its trace.

\appendix
\renewcommand{\thesection}{Appendix \Alph{section}}
\section{Moment hierarchy}
\label{app:hierarchy}
\renewcommand{\thesection}{\Alph{section}}

Multiplying Eq.~(\ref{eq:u_rec}) by $\sigma_i^{2n}$ and summing over $i$ gives
\begin{equation}
\label{eq:Fn_diff}
F_n(t+1) - F_n(t) = 4\lambda^2\, \nu\, \Omega_{n+1}
 - 4\lambda \bigl[ F_{n+1}(t) - \lambda\, \Omega_{n+1}\, F_1(t) - 2\lambda\, F_{n+2}(t) \bigr] ,
\end{equation}
an infinite hierarchy in which the equation for $F_n$ involves $F_{n+1}$ and
$F_{n+2}$. In the isotropic case all the $F_n$ are equal and the hierarchy
reduces to a single equation,
\begin{equation}
F_0(t+1) = \bigl[1-4\lambda+4\lambda^2(N+2)\bigr] F_0(t) + 4\lambda^2\nu N ,
\end{equation}
from which Eqs.~(\ref{eq:cond_equal}) and (\ref{eq:q}) follow. In the
stationary state, Eq.~(\ref{eq:Fn_diff}) becomes $F_{n}(\infty)-2\lambda
F_{n+1}(\infty)=\lambda\,\Omega_{n}[\nu+F_1(\infty)]$, which is solved by
$F_n(\infty)=\lambda[\nu+F_1(\infty)]\sum_{k\geq 0}(2\lambda)^k\Omega_{n+k}
=\lambda[\nu+F_1(\infty)]\Sigma_n(\lambda)$ and gives back
Eq.~(\ref{eq:u_inf}).

For the time-dependent problem, however, the hierarchy is not a convenient
starting point. The moments are not independent: if the $\sigma_i$ take $K\leq
N$ distinct values, the $F_n$ are linear combinations of $K$ quantities, and
$F_0,\ldots,F_{K-1}$ determine all the others (the corresponding matrix is of
Vandermonde type). Truncating the hierarchy at an arbitrary order with an
approximate closure relation therefore leads to ill-conditioned or singular
matrices. Moreover, the infinite matrix that propagates the hierarchy has a
spectrum that is unrelated to the dynamics. Indeed, the homogeneous part of
Eq.~(\ref{eq:Fn_diff}) without the term in $F_1$ acts on a sequence of the form
$F_n=z^n$ as the multiplication by $1-4\lambda z+8\lambda^2z^2$. The physical
sequences are the combinations of $z=\sigma_i^2\in(0,1]$, for which this factor
is $\rho_i$; bounded sequences with $|z|=1$ and $z\ne 1$, such as
$z=e^{i\theta}$, are also formal eigenvectors, and some of the corresponding
eigenvalues have a modulus larger than $1$ for any $\lambda>0$ (e.g.,
$1+4\lambda+8\lambda^2$ for $z=-1$), but they do not correspond to any set of
weight errors. Working with the $N$ variables $u_i(t)$, as in
Section~\ref{sec:dynamics}, removes these spurious modes.


\end{document}